# Imaging a Single-Electron Quantum Dot

*Parisa Fallahi,[1]† Ania C. Bleszynski,[1]† Robert M. Westervelt,[1*] Jian Huang,[1] Jamie D. Walls,[1] Eric J. Heller,[1] Micah Hanson,[2] Arthur C. Gossard[2]*

[1]Division of Engineering and Applied Sciences, Department of Physics, and Department of Chemistry and Chemical Biology, Harvard University, Cambridge, Massachusetts 02138, USA.

[2]Materials Department, University of California, Santa Barbara, California 93106, USA.

†These authors contributed equally to this work.

*Corresponding author. E-mail: westervelt@deas.harvard.edu.

Images of a single-electron quantum dot were obtained in the Coulomb blockade regime at liquid He temperatures using a cooled scanning probe microscope (SPM). The charged SPM tip shifts the lowest energy level in the dot and creates a ring in the image corresponding to a peak in the Coulomb-blockade conductance. Fits to the lineshape of the ring determine the tip-induced shift of the electron energy state in the dot. SPM manipulation of electrons in quantum dots promises to be useful in understanding, building and manipulating circuits for quantum information processing.

A scanning probe microscope (SPM) has proven to be a powerful tool to study, image, and manipulate mesoscopic systems. Electron waves in an open two-dimensional electron gas (2DEG) inside a semiconductor heterostructure were imaged using a liquid-He cooled SPM.[1-4] In the quantum Hall regime, edge states and localized states were seen.[5-9] Furthermore an SPM was used to study electrons confined in nanostructures; charge oscillations due to the Coulomb blockade in quantum dots formed in a carbon nanotube were observed.[10] Imaging a single electron spin has been accomplished recently using a magnetic resonance force microscope (MRFM), where the signal from a single electron's spin resonance (ESR) was detected.[11] Molecule cascades arranged on a clean surface in ultrahigh vacuum by a scanning tunneling microscope (STM) were used to perform logic operations.[12]

Single-electron quantum dots are promising candidates for quantum information processing. The electron spin in each dot acts as a qubit, and tunneling is used to entangle spins on adjacent dots.[13] To pursue these ideas, quantum dots that contain only one electron are being developed, as individual single-electron dots[14-18] and as tunnel-coupled single-electron dots.[19, 20] A useful circuit for quantum information processing will consist of many coupled quantum dots. Scanning probe microscopy promises to be important for the development and understanding of quantum dots and dot circuits, by providing ways to image electrons and to probe individual dots using electromagnetic fields.

In this letter, we show how a liquid-He cooled scanning probe microscope with a charged tip can image a single-electron quantum dot in the Coulomb blockade regime. The dot was formed in a GaAs/AlGaAs heterostructure by surface gates. The charged tip shifts the lowest energy level in the dot and creates a ring in the image corresponding to a peak in the Coulomb-blockade conductance. Fits to the lineshape of the ring determine the tip-induced shift of the electron energy state in the dot.

Figure 1 illustrates the imaging technique. A charged SPM tip scanned above the surface can change the induced charge in a quantum dot (Figure 1b) and change the number of electrons. An image is obtained by recording the dot conductance $G$ as the tip is scanned across the sample. The voltage $V_{tip}$ applied between the tip and the 2DEG perturbs the "bathtub" potential that holds electrons in the dot (Figure 1c). When the distance between the tip and the 2DEG is greater than the width of the wavefunction, as it was for the images in this paper, $V_{tip}$ moves the bathtub up and down without changing the wavefunction's shape significantly, in a manner similar to the side-gate voltage $V_G$. When the tip is sufficiently close to the dot, closer than the width of the wavefunction, the tip voltage $V_{tip}$ can change the shape of the wavefunction and thus the energy of the electron state above the bottom of the bathtub. The total shift $\Delta_{tip}$ in the ground state energy from both processes, indicated in Figure 1c, moves the gate-voltage position of the Coulomb-blockade conductance peak.

The quantum dot (Figure 1b) was formed in a $GaAs/Al_{0.3}Ga_{0.7}As$ heterostructure by Cr surface gates. The heterostructure contains a 2DEG 52nm below the surface, with measured density $3.8 \times 10^{11}$ cm$^{-2}$ and mobility 470,000 cm$^2$V$^{-1}$s$^{-1}$ at 4.2 K. The heterostructure was grown by molecular beam epitaxy with the following layers: 5 nm GaAs cap layer, 25 nm $Al_{0.3}Ga_{0.7}As$, Si delta-doping layer, 22 nm $Al_{0.3}Ga_{0.7}As$, 20 nm GaAs, 100 nm $Al_{0.3}Ga_{0.7}As$, a 200 period $GaAs/Al_{0.3}Ga_{0.7}As$ superlattice, 300 nm GaAs buffer and a semi-insulating GaAs substrate. The 2DEG is formed in a 20 nm wide GaAs square well between two $Al_{0.3}Ga_{0.7}As$ barriers. The sample was mounted in a liquid-He cooled SPM[14,15] and cooled to T = 1.7 K.

Without the tip present, the quantum dot could be tuned to contain 0 or 1 electrons in the Coulomb blockade regime. This is clearly shown in Figure 2a, which plots the differential dot conductance $dI/dV_{SD}$ vs. source-to-drain voltage $V_{SD}$ and side gate voltage $V_G$ at T = 1.7 K. The conductance peaks correspond to resonant tunneling through a single quantum state. Coulomb blockade diamond measurements reveal an appreciable amount of information: from Figure 2a we determine the one-electron charging energy 4.2 meV and the ground-state to first-excited-state energy spacing 3.1 meV.

Figure 2b shows the differential conductance $g = dI/dV_{SD}$ vs. source-to-drain voltage $V_{SD}$ for fixed $V_G$; here the tip voltage $V_{tip}$ is used to change the induced charge on the dot with the tip held at a fixed position near the dot. The pattern of Coulomb blockade diamonds is similar to those obtained by varying $V_G$ in Figure 2a, demonstrating that the tip acts as a gate, and that the tip to dot coupling is similar to the coupling between the side-gate and the dot. Because the tip can be arbitrarily positioned over the sample, it can be used as a movable gate to change the number of electrons on a quantum dot, as well as to direct electrons in a desired direction. These abilities promise to be very useful for the development of quantum dot circuits for quantum information processing.

Images of the single-electron quantum dot were obtained at T = 1.7 K by recording the Coulomb blockade conductance with $V_{tip}$ and $V_G$ fixed, and $V_{SD} = 0$ V, while the tip was spatially scanned over the quantum dot, 100 nm above the surface. A series of images are shown in Figures 3a-d for tip voltages 40mV, 50mV, 60mV and 80mV respectively. The field of view covers an area within the gates of the quantum dot (see Figure 1b). In each image, a ring-shaped feature is observed, centered on the middle of the dot. The ring represents a contour of constant tip to dot coupling at which the Coulomb blockade conductance is on a peak. This peak corresponds to resonant tunneling through the lowest energy level of the dot. The dot contains one electron when the tip is outside the ring and zero electrons when the tip is inside the ring. To confirm that the dot is empty inside the ring, we moved $V_G$ to more-negative voltages and verified that no additional conductance peaks appeared.

The strength of the interaction between the SPM tip and the dot can be adjusted by changing the tip voltage $V_{tip}$ as shown in Figures 3a-d. In Figure 3a the tip pushes the electron off the dot when the tip is about 100 nm to the side of the center. As $V_{tip}$ is increased in a series of steps from Figures 3a to 3d, the radius of the ring shrinks to a small value. As discussed below, the lineshape of the ring provides a window through which one can extract information about the dot. The probing window can be moved to any desired location with respect to the dot by changing the ring radius.

The spatial resolution in Figures 3a-d is quite good, finer than the width of the tip electrostatic potential $\Phi_{tip}(\vec{r},\vec{r}_e)$ at the point $\vec{r}_e$ in the 2DEG for tip position $\vec{r}$; this width is determined in part by the height of the tip above the surface. The resolution is enhanced by the strong dependence of the Coulomb blockade conductance $G$ on the change $\Delta_{tip}$ in electron energy. However, the images in Figure 3 do not determine the shape of wavefunction amplitude $|\psi(\vec{r}_e)|^2$, because it is much narrower than $\Phi_{tip}(\vec{r},\vec{r}_e)$ for this case.

Simulated images of the single-electron quantum dot are shown in Figures 3e-h. These images show the calculated dot conductance as a function of lateral tip position using parameters from the experiment, including the tip voltage and height. In these calculations the dot was assumed to have a parabolic confining potential with an energy level spacing $\Delta E = 3.1$ meV matching the measured value for the first excited state from Figure 2a. The ground state energy of the dot in the presence of the tip was obtained by solving Schrödinger's equation for this system. The dot conductance was calculated in the resonant tunneling regime, involving only a single energy level in the dot.[21] The simulations in Figures 3e-h show rings of high dot conductance that are in good agreement with the experimental images (Figures 3a-d). Changes in ring diameter with changing tip voltage accurately match the experimental images.

Maps of the tip induced shift in energy level vs. tip position obtained from the lineshape of the Coulomb blockade rings in Figures 3a-d are shown in Figures 3i-l. These maps were determined in the following way. For resonant tunneling, the lineshape is given by[21]

$$G = G_{max}\left[Cosh(\Delta/2k_BT)\right]^{-2}, \tag{1}$$

where $\Delta$ is the energy difference between the lowest energy level in the dot and the Fermi energy in the leads. The energy difference is zero at resonance and deviates from zero as the tip shifts the energy level upwards or downwards. The dot conductance at resonance is:[21]

$$G_{max}(\vec{r}) = \left(e^2/4k_BT\right)\Gamma(\vec{r}). \tag{2}$$

The tunneling rate $\Gamma(\vec{r})$ alters as the tip is scanned above the dot, due to changes in the coupling between the tip and the point contacts, resulting in variations in $G_{max}$ along the ring as seen in Figures 3a-d. The values of $G_{max}(\vec{r})$ in Eq. 2 used to compute the maps were obtained from a smooth two-dimensional polynomial function that was fit to the measured values of $G_{max}$ along the crest of the ring. The strong dependence of the Coulomb blockade conductance on $\Delta$ allows us to measure the energy shift accurately.

If the SPM tip is sufficiently close to the 2DEG, at distances less that the width of the electron wavefunction $|\psi(\vec{r}_e)|^2$, it is theoretically possible to extract the shape of the wavefunction inside the dot from SPM images.[22] The wavefunction $|\psi(\vec{r}_e)|^2$ can be extracted from a map of the dot energy level shift $\Delta(\vec{r})$, where $\Delta(\vec{r})$ equals $\Delta_{tip}(\vec{r})$ plus a constant determined by the side-gate voltage $V_G$. The tip voltage $V_{tip}$ is adjusted to produce only a weak tip perturbation $\Phi_{tip}(\vec{r},\vec{r}_e)$, the change in electrostatic potential due to the tip in the plane of the 2DEG. From first-order perturbation theory, $\Delta_{tip}(\vec{r})$ is the convolution of the wave function of the electron in the dot and the tip potential:

$$\Delta_{tip}(\vec{r}) = \langle \psi | \Phi_{tip}(\vec{r},\vec{r}_e) | \psi \rangle = conv\left(|\psi(\vec{r}_e)|^2, \Phi_{tip}(\vec{r},\vec{r}_e)\right). \tag{3}$$

Knowing the shape of $\Phi_{tip}(\vec{r},\vec{r}_e)$, one can deconvolve measurements of $\Delta_{tip}(\vec{r})$ using Eq. 3 to extract the shape of the unperturbed wavefunction amplitude $|\psi(\vec{r}_e)|^2$. For the images presented in this paper, the tip perturbation was wider than the wavefunction, and this method is not applicable. In future experiments we hope to extract the shape of the wavefunction using a relatively narrow tip perturbation.

ACKNOWLEDGEMENTS

This work was supported at Harvard University by DARPA grant DAAD19-01-1-0659 and by the Nanoscale Science and Engineering Center (NSEC) under NSF grant PHY-0117795; work at UC Santa Barbara was supported by the Institute for Quantum Engineering, Science and Technology (iQUEST).

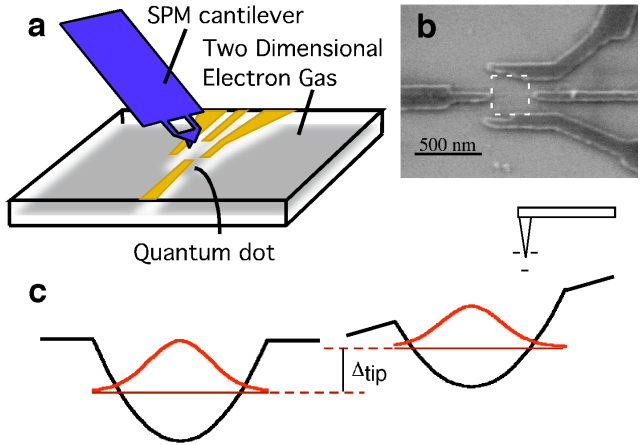

**Figure 1.** (a) Schematic diagram of the experimental set-up used to image electrons in a single-electron quantum dot. A charged scanning probe microscope (SPM) tip is scanned at a fixed height above the surface of the GaAs/AlGaAs heterostructure containing the dot. Images are obtained by recording the Coulomb blockade conductance G vs. tip position. (b) A scanning electron micrograph of the quantum dot. The dashed line indicates the area covered by the conductance images. (c) Schematic diagram that shows how the potential holding an electron in the dot is affected by the charged tip. The tip-induced shift $\Delta_{tip}$ in the energy of the electron state changes the Coulomb blockade conductance of the dot.

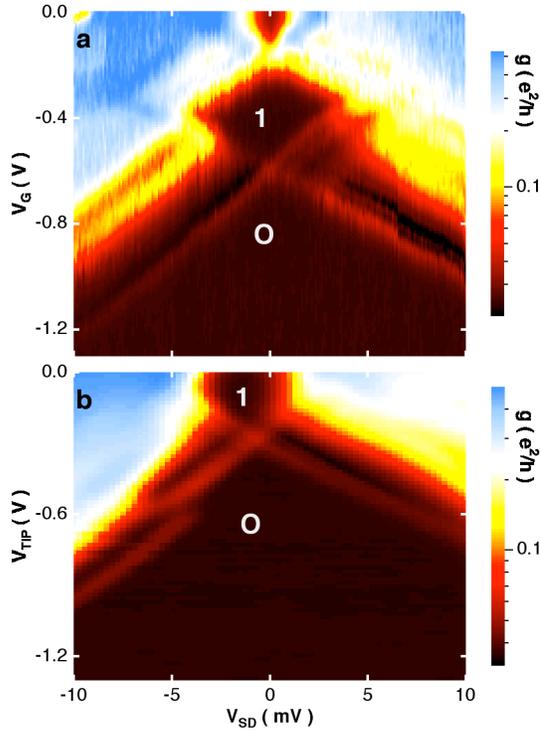

**Figure 2.** (a) Plot of differential conductance $g = dI/dV_{SD}$ as a function of side-gate voltage $V_G$ and source-to-drain voltage $V_{SD}$ at $T = 1.7$ K, showing Coulomb blockade diamonds for 0 and 1 electrons, and resonant tunneling through the ground and first excited energy levels separated by 3.1 meV. (b) Plot of differential conductance $g = dI/dV_{SD}$ vs. SPM tip voltage $V_{tip}$ and $V_{SD}$ at $T = 1.7$ K for a fixed tip position, showing that the tip acts as a moveable gate.

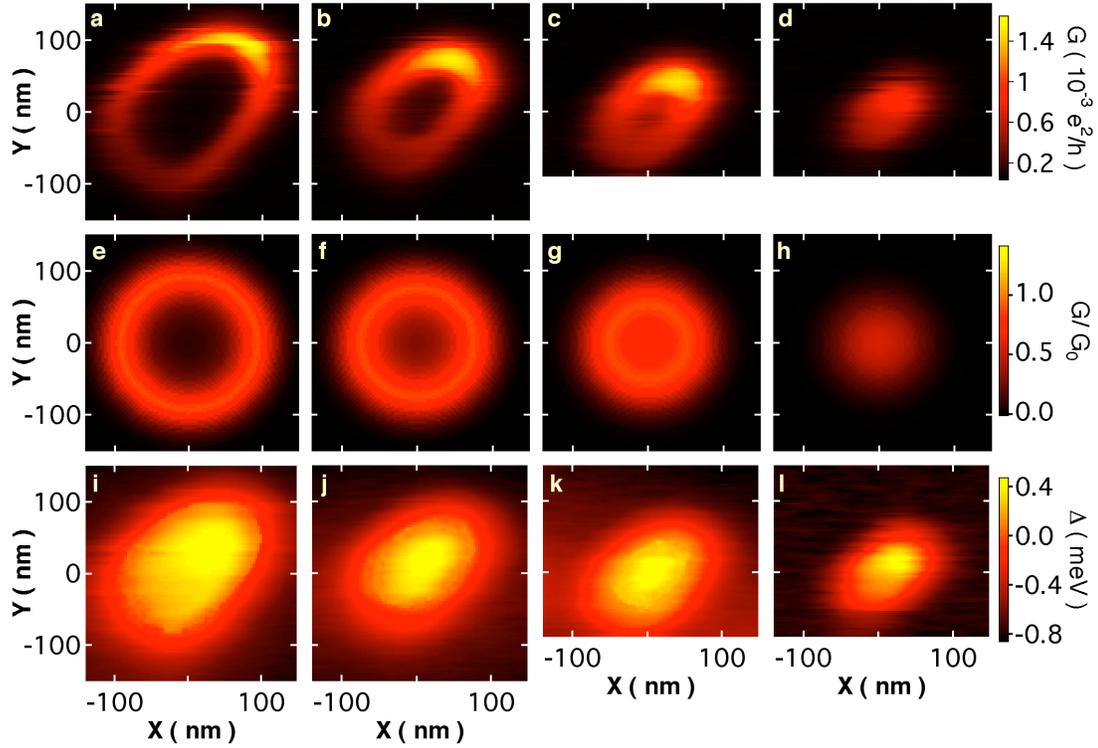

**Figure 3.** (a-d) Coulomb blockade images of a single-electron quantum dot at $T = 1.7$ K, showing the dot conductance G vs. tip position. The ring of high conductance around the center of the dot is formed by the Coulomb blockade peak between 0 and 1 electron in the dot. The tip voltages $V_{tip}$ for A-D are 40 mV, 50 mV, 60 mV and 80 mV respectively. (e-h) Theoretical simulations of the images in a-d for a dot formed by a parabolic potential with energy spacing 3.1 meV (energy of first excited state from Figure 2a) for the same tip voltages as a-d. (i-l) Experimental maps of the energy shift Δ of electrons in the dot vs. tip position, extracted from the measured lineshape of the Coulomb blockade conductance peak forming the rings in images a-d.

REFERENCES

(1) Topinka, M. A.; Westervelt, R. M.; Heller, E. J. *Phys. Today* **2003**, December, 47.

(2) Topinka, M. A.; LeRoy, B. J.; Shaw, S. E. J.; Heller, E. J.; Westervelt, R. M.; Maranowski, K. D.; Gossard, A. C. *Science* **2000**, 289, 2323.

(3) Topinka, M.A.; LeRoy, B. J.; Westervelt, R. M.; Shaw, S. E. J.; Fleischmann, R.; Heller, E. J.; Maranowski, K. D.; Gossard, A. C. *Nature* **2001**, 410, 183.

(4) Crook, R.; Graham, A. C.; Smith, C. G.; Farrer, I.; Beere, H. E.; Ritchie, D. A. *Nature* **2003**, 424, 751.

(5) Finkelstein, G.; Glicofridis, P. I.; Ashoori, R. C.; Shayegan, M. *Science* **2000**, 289, 90.

(6) Yacoby, A.; Hess, H. F.; Fulton, T. A.; Pfeiffer, L. N.; West, K. W. *Solid State Commun.* **1999**, 111, 1.

(7) Zhitenev, N. B.; Fulton, T. A.; Yacoby, A.; Hess, H. F.; Pfeiffer, L. N.; West, K. W. *Nature* **2000**, 404, 473.

(8) Ahlswede, A.; Weitz, P.; Weis, J.; von Klitzing, K.; Eberl, K. *Physica B* **2001**, 298, 562.

(9) Ihn, T.; Rychen, J.; Vancura, T.; Ensslin, K.; Wegscheider, W.; Bichler, M. *Physica E* **2002**, 13, 671.

(10) Woodside, M. T.; McEuen, P. L. *Science* **2002**, 296, 1098.

(11) Rugar, D.; Budakian, R.; Mamin, H. J.; Chui, B. W. *Nature* **2004**, 430, 329.

(12) Heinrich, A. J.; Lutz, C. P.; Gupta, J. A.; Eigler, D. M. *Science* **2002**, 298, 1381.

(13) Loss, D.; DiVincenzo, D. P. *Phys. Rev. A* **1998**, 57, 120.

(14) Kouwenhoven, L. P.; Austing, D. G.; Tarucha, S. *Rep. Prog. Phys.* **2001**, 64, 701.

(15) Tarucha, S.; Austing, D. G.; Honda, T.; van der Hage, R. J.; Kouwenhoven, L. P. *Phys. Rev. Lett.* **1996**, 77, 3613.

(16) Ashoori, R. C. *Nature* **1996**, 379, 413.

(17) Ciorga, M.; Sachrajda, A. S.; Hawrylak, P.; Gould, C.; Zawadzki, P.; Jullian, S.; Feng, Y.; Wasilewski, Z. *Phys. Rev. B* **2000**, 61, R16315.

(18) Potok, R. M.; Folk, J. A.; Marcus, C. M.; Umansky, V.; Hanson, M.; Gossard, A. C. *Phys. Rev. Lett.* **2003**, 91, 016802-1.

(19) Elzerman, J. M.; Hanson, R.; Greidanus, J. S.; Willems van Beveren, L. H.; De Franceschi, S.; Vandersypen, L. M. K.; Tarucha, S.; Kouwenhoven, L. P. *Phys. Rev. B* **2003**, 67, 161308-1.

(20) Chan, I. H.; Fallahi, P.; Vidan, A.; Westervelt, R. M.; Hanson, M.; Gossard, A. C. *Nanotechnology* **2004**, 15, 609.

(21) Beenakker, C. W. J. *Phys. Rev. B* **1991**, 44, 1646.

(22) Fallahi, P.; Bleszynski, A. C.; Westervelt, R. M.; Huang, J.; Walls, J. D.; Heller, E. J.; Hanson, M.; Gossard, A. C. *Proc. 27 Int. Conf. on Physics and Semiconductors (ICPS27), Flagstaff, July 26-30, 2004,* in press.